\documentstyle[12pt,aas2pp4]{article}
\begin{document} 

\lefthead{Mazzali et al.}
\righthead{Nebular velocities in SNe Ia and their relationship to light curves}

\title{Nebular velocities in type Ia Supernovae and their relationship to light curves
 \footnote[0]{Based on observations collected at ESO-La Silla (Chile).} }

\author{P.A. Mazzali\altaffilmark{1}, E. Cappellaro\altaffilmark{2},
I.J. Danziger\altaffilmark{1}, M. Turatto\altaffilmark{2}, 
S. Benetti\altaffilmark{3} }

\altaffiltext{1}{Osservatorio Astronomico di Trieste, via G.B. Tiepolo 11, 
I-34131 Trieste, Italy}
\altaffiltext{2}{Osservatorio Astronomico di Padova, vicolo 
dell'Osservatorio 5, I-35122 Padova, Italy}
\altaffiltext{3}{European Southern Observatory, Alonso de Cordova 3107,
Vitacura, Casilla 19001 Santiago 19, Chile}

\begin{abstract}

Recently gathered observational data on a sample of Type Ia Supernovae
(SNe~Ia) reveal a wide distribution of expansion velocities of the Fe
cores, measured from the width of the nebular lines. Moreover, the
velocity appears to correlate with the luminosity decline rate
after maximum light, $\Delta m_{15}(B)$. Since it has been shown that
for SNe~Ia $\Delta m_{15}(B)$ correlates with the absolute magnitude
at maximum, this then implies a relation between the expansion
velocity of the Fe nebula and the luminosity at maximum.

Physically, the maximum luminosity is related to the mass of
synthesized $^{56}$Ni, whereas the $FWHM$ of the lines is related to
the kinetic energy of the ejecta. Our finding constitutes
observational proof of the theoretical prediction that the two
quantities have to be related.
\end{abstract}


\section {Introduction}

Type Ia Supernovae (SNe~Ia) are thought to result from the explosion of
mass-accreting white dwarfs (WD). Since SNe~Ia are bright, they can be used to
determine distances if their absolute brightness can be calibrated. The fact
that all SNe~Ia seemed to show similar light curves and spectra encouraged
their use as standard candles. In addition, SNe~Ia are important for galaxy
chemical evolution since they provide most of the Fe in the universe.

With the accumulation of new and more accurate data, the paradigm that SNe~Ia
are all similar has been questioned. In particular, the discovery of objects
such as the very bright SN~1991T (Filippenko et al. 1992a; Phillips et al.
1992; Ruiz-Lapuente et al. 1992) and the very
faint SN~1991bg (Filippenko et al. 1992b; Leibundgut et al. 1993; Turatto et
al. 1997) showed that SNe~Ia exist with very different properties from those of
the classical objects defining this class, e.g. SNe 1981B (Branch et al. 1983)
or 1992A (Kirshner et al. 1993).
Other peculiar SNe~Ia, both over- and under-luminous, have subsequently been
discovered, showing that such events, although less frequent than `normal' 
SNe~Ia, are certainly not just rare exceptions. The progenitor masses and the
explosion mechanisms leading to these peculiar SNe~Ia are currently topics of
debate, but it is clear that the over-luminous objects produce more $^{56}$Ni
than `normal' SNe~Ia, while the under-luminous ones produce less (Mazzali et
al. 1993, 1995, 1997).

Objects such as SNe~1991bg and 1991T are not only different in luminosity from
the average SN~Ia, but their spectra are also different, at least at phases
before and around maximum (however, Phillips et al. (1992) have shown that 
SN~1991T 14 days after maximum was indistinguishable from other SNe~Ia), 
reflecting differences in temperature and also in the composition of the outer
layers.  It has been argued that  by selecting SNe Ia on the basis of a
'normal' spectrum at about maximum or earlier, one can define a subgroup of
`normal' SNe~Ia which can be used as standard candles (Branch \& Miller 1993;
Vaughan et al. 1995; Saha et al. 1997). The vast majority of well-observed
SNe~Ia belong to this group, and so peculiar objects such as SNe~1991bg and
1991T can be safely removed from the sample.

Therefore, if the `spectroscopically normal' SNe~Ia are to be used as standard
candles, all of their properties should be similar. However, observational
evidence has been accumulating for the existence of a relation between the
absolute magnitude at maximum of SNe~Ia (normal and peculiar) and the rate at
which their light curves decline after maximum.  This is known as the Phillips
or $M - \Delta m_{15}$ relation, which indicates that the SNe whose rate of
decline is slower are those that reach a brighter absolute magnitude (Phillips
1993).  Indeed, the Hubble diagram for a sample of distant SNe~Ia has a much
smaller dispersion when the $M - \Delta m_{15}$ effect is corrected for (Hamuy
et al. 1996b).

Model light curves have shown that the origin of the $M_B - \Delta m_{15}$
relation could lie in the existence of a range in the Ni mass and distribution
produced in the explosion (H\"oflich et al. 1996; Cappellaro et al. 1997).  The
amount of Ni produced is directly related to the energy of the explosion and 
is reflected in the kinetic energy of the ejecta. Thus we address the question:
can any direct observational evidence for this scenario be found?

\section {The late-time spectra}

We have analyzed the photometric and spectroscopic data on SNe~Ia collected at
ESO over the last several years. We concentrated on spectra obtained at epochs
of about one year after maximum. At epochs greater then 100-150 d, SNe~Ia
display a nebular spectrum: the ejecta have become so thin that they are
transparent to radiation.  The gas is heated by the deposition of some of the
energy carried by the $\gamma$-rays and the positrons produced by the decay 
of $^{56}$Co (the daughter nucleus of $^{56}$Ni) into stable $^{56}$Fe, while
cooling takes place via forbidden line emission. Forbidden lines of Fe~II and 
Fe~III dominate the spectrum, confirming that iron-group elements are prevalent 
in the inner part of the ejecta. The outer layers, where lighter elements 
are distributed, have too small a column density to absorb the surviving
$\gamma$-rays (the positrons, which actually dominate the energy input after
about 200 days, deposit essentially in situ) and thus do not contribute
significantly to the spectrum.

A representative sample of SNe Ia spectra at an epoch of about 1  year taken
from the ESO+Asiago SN archive is shown in Fig.1.  Although the spectra shown
in Fig.1 belong to SNe that differ by up to 3 magnitudes in brightness at
maximum (from SN~1991bg to SN~1991T), it is interesting to note that their
nebular spectra are quite similar, except  for the line widths. In particular,
the two strongest lines in the blue have similar ratios, when reddening has
been corrected for. Since the bluer feature (4700\AA) is almost exclusively 
due to Fe~III forbidden lines, while the red feature (5300\AA) is formed by 
a blend of Fe~II and Fe~III forbidden lines, their ratio is sensitive to the
ionization/excitation conditions (and hence temperature and density in the
nebula). Preliminary modelling work on the nebular spectra (Mazzali et al.
1998, in prep.) indicates that the observed range of a factor 1.5 in the line
ratio can be obtained for a difference in temperature and ionization of not
more than 10\%. This must imply that the energy deposition rate per unit volume
in the SN nebula is quite similar in all SNe.  Indeed Cappellaro et al. (1997)
showed that the difference in brightness at maximum is preserved, or even
increased  at the later epochs.  Since our sample includes objects from the
least luminous SN~Ia known to date, SN~1991bg, to the most luminous one,
SN~1991T, the entire range of SNe~Ia appears to share these same properties.

The composition of the emitting nebula appears to be similar in all SNe: all
spectra are dominated by lines of Fe and Co, although the mass of these
elements produced is probably very different in the various objects, ranging
from about 0.1 M$_{\odot}$ in SN~1991bg (Mazzali et al. 1997) to approximately 
0.6 M$_{\odot}$ in `normal' SNe~Ia and about 1 M$_{\odot}$ in SN~1991T 
(Spyromilio et al. 1993).  This confirms that only 
the inner Fe sphere is significantly affected by the heat input following the
absorption of the $\gamma$-rays and the positrons, essentially a local process,
and that mixing is not significant, at least at the lowest velocities.

The second distinguishing property of the nebular spectra of Fig.1 is that 
the lines have different widths in different SNe.  The line widths span about 
a factor of 5: the faint SN~1991bg has the narrowest lines, and the bright
SN~1991T has the broadest lines. 

If the physical conditions (temperature, density) in the nebulae are
similar, as indicated by the line ratios, and if the brighter SNe
produce more $^{56}$Ni, it should be expected that the more massive Ni
sphere in the brighter and more energetic SNe is larger than in the
fainter SNe. Since the expansion of the ejecta is homologous (i.e. $v
\propto r$), a correlation between the width of the nebular lines and
the brightness of the SN should be expected. 

Excluding the case of SN~1991bg, where several individual lines are
resolved, in all other cases the observed lines are the result of very
strong blending.  In particular, the features in the range
4500-5500\AA\ can be fitted with three gaussians to measure position,
intensity and FWHM (the continuum being set to zero).  The FWHM of the
strongest feature, measured at $\sim4700$\AA, was then selected as the
reference value, because it is always narrower than that at $\sim5300$\AA\ 
and has a much better S/N than the much weaker feature at $\sim5000$\AA.

The value of FWHM is representative of the ejecta velocity, but a
knowledge of the density/velocity distribution would be needed to
translate it into a kinetic energy. In addition, when blending is severe, 
the value of FWHM is an overestimate of the velocity of the Fe nebula.

The effect of line blending can be estimated using synthetic spectra
(Mazzali et al. 1998, in prep.). By comparing the velocity used in the
model for the Fe-nebula with the FWHM of the $\sim4700$\AA\ feature
measured on the model synthetic spectra we found that blending can be
neglected only for input velocities of up to 3000 km/s.  The relation
between input velocity and measured FWHM has a discontinuity when
blending sets in, between about 3000 and 6000 km/s, but for higher
velocities a monotonic relation is recovered. The velocity of the
model Fe-nebula is marked in the upper x-axis of Fig.~2.

The measured FWHM should be plotted against the SN absolute magnitude. 
Unfortunately, the absolute magnitude is a very uncertain quantity, since it
depends on the adopted distance scale and extinction correction.  
To avoid these problems, we make use of the relation between the absolute 
magnitude at maximum and $\Delta m_{15}(B)$, the difference in $B$ magnitude 
between the time of maximum and 15 days later (Phillips 1993, 
Hamuy et al. 1996b). This latter quantity can be measured directly, 
so that two observables can be compared. 
Measurements of $\Delta m_{15}(B)$ are readily available in the literature 
for most of the objects in our sample.  For SNe 1994ae and 1995D, we measured
$\Delta m_{15}(B)$ on the light curves published by Riess et al. (1996),
whereas for SNe 1991M and 1993L we used unpublished ESO photometry, which
unfortunately does not cover the time of maximum very well.  
We also note that the light curve of SN~1989M is very poorly sampled. 

A plot of FWHM(4700\AA) v. $\Delta m_{15}(B)$ is shown in Figure 2.  It is
apparent that a correlation exists between the two quantities (the Spearman
rank-order correlation coefficent is $-0.81$). Note also that a significant 
spread in both line width and decline time is present even among `normal'
SNe~Ia.  Even when the extreme case of SNe 1991bg is removed from the sample,
the null hypothesis of no correlation between $\Delta m_{15}(B)$ and
FWHM(4700\AA) has a probability of only 0.003. Furthermore, the FWHM measured
for SN~1991bg is lower than for other SNe both because of the intrinsically
low velocity and the absence of blending.

\section {Discussion}

Our results establish a connection between the luminosity of the SNe and their
kinetic energy. Since the maximum luminosity of a SN~Ia depends essentially on
the mass of $^{56}$Ni produced, the observed difference in line width and 
strength in the late-time spectra is a strong indication that SNe~Ia produce 
a range of Ni masses.  In the case of the peculiar SNe~Ia, the properties at 
early times are sufficiently different that the spectra are affected, while 
for `spectroscopically normal' SNe this is not the case, and the differences
emerge only in the rate of decline and in the nebular line width. However,
Branch \& van den Bergh (1993) noted a sequence in the velocity of the Si II
lines in SNe~Ia about 10 days after maximum. This relation to galaxy type may
be a less quantitative manifestation of the same phenomenon we are witnessing
at late time. Also, Nugent et al. (1995) suggested a `spectroscopic sequence'
among SNe~Ia near maximum. All the SNe in their sample are also included in
ours, and the ordering of the objects by nebular line width is similar to what 
they suggest based on the value of $V(Max)$. They also hypothesized that the 
sequence is the result of the different amounts of $^{56}$Ni produced by the 
various SNe. 

We can assume that the line width defines the volume of the Ni sphere, and that
this translates directly into a Ni mass. If we accept the estimate of the Ni
mass for SN 1991T ($\sim 1.0$ M$_{\odot}$, Spyromilio et al. 1993), we can
scale the other SNe accordingly.  We find then that SN 1986G should have a Ni
mass of $\sim 0.3$ M$_{\odot}$; objects such as SNe 1981B and 1992A, which we
could call `faint normal', have $\sim 0.4-0.5$ M$_{\odot}$, while `bright
normal' objects like SN~1990N and 1994D have $\sim 0.5-0.6$ M$_{\odot}$, and
appear to be the only SNe producing the `canonical' amount of $^{56}$Ni.  
These values are only rough estimates, and further modelling work is required 
to obtain more accurate results. Note that Ruiz-Lapuente \& Lucy (1992)
determined for SN~1986G M($^{56}$Ni$) = 0.38 \pm 0.03$ M$_{\odot}$, but they
used a larger reddening. Finally, the very subluminous SN~1991bg appears to 
have a very small Ni mass, of about 0.01 M$_{\odot}$. This is smaller than 
the published values (Mazzali et al. 1997), but  
note that in the case of SN~1991bg blending does not affect the line width, 
and that the ratio of the fluxes in the blue and the red feature is the 
smallest among all the SNe in our sample, indicating a lower degree of 
ionization and therefore a smaller fractional abundance of the Fe III ion, 
which is responsible for the blue feature. 
The models for SN~1991bg (Mazzali et al. 1997) indeed require a higher Ni 
density than in other SNe. This leads to a lower ionization degree, and the  
flux is emitted more in the red line near 5300\AA. 

Models for the explosion of WD's leading to SNe~Ia which produce less than 
the canonical 0.6~M$_{\odot}$ of $^{56}$Ni typically require progenitors less
massive than the Chandrasekhar limit of 1.4~M$_{\odot}$ (see e.g. Livne \&
Arnett 1995 and references therein). H\"{o}flich et al. (1995) presented 
Chandrasekhar mass, pulsating delayed detonation models for SNe~Ia. They can
vary the mass of $^{56}$Ni produced by varying the density at which the
transition from deflagration to detonation occurs. Models that synthesize less 
$^{56}$Ni produce more Si-group elements, so that all models produce about the
same kinetic energy. They claim this family of models can explain the observed
differences among SNe~Ia light curves. Our results prove that the kinetic
energy is actually related to the $^{56}$Ni mass, so that burning to NSE is
necessary in all realistic explosion models. Since the rapid decline of the
faint SNe~Ia does not appear to be the result of a large kinetic energy in a 
Chandrasekhar-mass explosion, it is then likely to be due to a small exploding
mass.  The fact that SN~1991bg was the result 
of the explosion of a sub-Chandrasekhar WD has received support from several
independent studies.  If the range of $^{56}$Ni masses which we observe in 
the nebular spectra is indeed the result of a range of progenitor masses, our
understanding of the physics and the nature of SNe~Ia would be revolutionized.

Formally, we can fit the data of Fig.2 with a straight line. 
Since errors affect both axes, we performed a weighted, linear least
square fit (Press et al. 1992), which gives the line shown in Fig.2.
However, the variance of the points around the line seems too large to 
be due only to the estimated errors (the reduced $\chi^2 = 4.1$, 
and even excluding SN~1991bg we get $\chi^2 = 2.54$)

Whereas a possible interpretation of the large variance of the points in Fig.2
is that the relation $\Delta m_{15}(B)$ vs $FWHM_{300d}$ is not linear, it is
possible that we are seeing an indication of the existence of a `second
parameter' determining the shape of the light curves (or possibly of the 
velocities).

A similar suggestion was recently made by H\"{o}flich et al. (1996) in the
context of a discussion of the $M_V - \Delta m_{20}(V)$ relation. As the same
authors commented, however, the uncertainties in the distance moduli and in 
the extinction corrections, and hence in the calibration of the SN absolute
magnitude, cast some doubts on the reality of their finding. The variables
plotted in Fig.~2 are free from such uncertainties, and therefore the result is
particularly intriguing. This `second parameter' could be related to the mode
of explosion, and possibly also to the difference in the  masses of the
progenitors through the ratio of the mass of synthesized $^{56}$Ni and the mass
of the ejecta (Cappellaro et al. 1997).  The existence of a `second parameter'
might weaken the potential of using the decline rate as a means to determine SN
luminosities.

However, the nebular spectra may offer another possibility.  From a theoretical
point of view, the absolute magnitude at maximum should be directly correlated
to the mass of synthesized $^{56}$Ni, with an associated uncertainty 
depending on the explosion model.  If the $^{56}$Ni mass can be obtained from 
the nebular spectra, we may be able to use the late-time spectra of SNe~Ia as 
distance indicators, thus avoiding the uncertainties introduced by the `second
parameter'.  If a relation can be established between the width of the nebular
lines and the absolute magnitude at maximum, and once it is well calibrated
with the acquisition of more data, the absolute magnitude of a SN, and hence
its distance, could be determined straightforwardly from a line width
measurement on a nebular spectrum at an epoch of about one year.  Precise
timing of the spectra is not so important, since we do not expect and have not
seen any significant systematic evolution of the line width while the SNe are
in their nebular phase. Also, theoretical models can be used to derive the
$^{56}$Ni mass (Ruiz-Lapuente \& Lucy 1992) and hence the luminosity at
maximum (Cappellaro et al. 1997). Against this however is the fact that the 
SNe at one year are at least 7 magnitudes fainter than at maximum light.

If SNe~Ia produce a continuous range of $^{56}$Ni masses and kinetic energies, 
it is also quite likely that the masses of the progenitors vary, thus 
favouring the sub-Chandrasekhar scenario for a large number of SNe~Ia. Even 
objects that have been regarded as normal because of their spectroscopic 
appearance may indeed be subluminous and submassive, although less so than the 
well known cases, which are only the extremes of a continuous distribution.

\bigskip  
%{\bf Acknowledgements.} 
This work was conducted in the
framework of the ESO Key Programme: Supernovae. It is a pleasure to
thank L.B.Lucy for several useful duscussions and comments on an
earlier version of this manuscript.

\begin{deluxetable}{lcrrlcrlrrl}
\scriptsize
\tablenum{10}
\tablecaption{The sample of SNe Ia}
\tablehead{\colhead{SN} &
\colhead{epoch} &
\colhead{$V_{max}$} &
\colhead{$\Delta m_B(15)$} &
\colhead{$A_V$} &
\colhead{ref} &
\colhead{ $\mu$} &
\colhead{ref (meth)  } &
\colhead{ $M_V$ } &
\colhead{FWHM }&
\colhead{ref} \nl
\colhead{ }& {days} &
  & & & & & & & 
\colhead {(10$^3$ km s$^{-1}$)} & }
\startdata       				  	  	     
1981B  & 300 & 11.93 & $1.10\pm0.05$ & 0.00    & a  & 31.10&l (Ceph)    & $-19.17\pm0.15$& $14.1\pm0.5$ & n\nl
1986G  & 300 & 11.38 & $1.73\pm0.07$ & 1.95    & a  & 27.86&a (SBF/PNLF)& $-18.43\pm0.32$& $12.2\pm0.4$ & o \nl
1989B  & 330 & 11.95 & $1.31\pm0.07$ & 1.05    & b  & 30.28&l (Ceph)    & $-19.38\pm0.18$& $12.6\pm1.9$ & p \nl
1989M  & 300 & 12.30 & $1.10\pm0.20$ & 0.11\tablenotemark{*}& c  & 31.10&c (T-F)     & $-18.91\pm0.36$& $17.7\pm0.4$ & E\nl
1990N  & 245 & 12.70 & $1.03\pm0.06$ & 0.00    & d,a& 32.03&l (Ceph)    & $-19.33\pm0.23$& $15.4\pm0.8$ & q \nl
1991M  & 150 & 14.10 & $1.50\pm0.10$ & 0.05\tablenotemark{*}& E  & 32.72&$\#$  (v3K) & $-18.67\pm0.60$& $14.0\pm0.7$ & E\nl
1991T  & 282 & 11.51 & $0.95\pm0.05$ & 0.39    & d,m& 30.60&b (T-F)     & $-19.48\pm0.30$& $18.0\pm0.5$ & m\nl
1991bg & 203 & 13.96 & $1.95\pm0.10$ & 0.15    & f  & 31.65&i (GCLF)    & $-17.84\pm0.36$& $ 3.5\pm1.0$ & f\nl
1992A  & 300 & 12.55 & $1.47\pm0.05$ & 0.00    & a  & 31.35&i (GCLF)    & $-18.79\pm0.16$& $14.8\pm0.4$ & E \nl
1993L  & 269 & 13.20 & $1.50\pm0.10$ & 0.03\tablenotemark{*}& E  & 31.64&$\#$  (v3K) & $-18.47\pm0.60$& $13.3\pm0.6$ & E\nl
1994D  & 300 & 11.90 & $1.26\pm0.05$ & 0.09    & g,h& 30.86&a (SBF)     & $-19.05\pm0.12$& $15.3\pm0.4$ & E \nl
1994ae & 182 & 13.05 & $1.00\pm0.10$ & 0.42    & e  & 31.28&e (T-F)     & $-18.65\pm0.41$& $14.6\pm0.6$ & E \nl
1995D  & 363 & 13.40 & $1.00\pm0.10$ & 0.21    & e  & 32.73&$\#$  (v3K) & $-19.54\pm0.40$& $15.2\pm1.0$ & E\nl
1996X  & 299 & 13.25 & $1.30\pm0.10$ & 0.20\tablenotemark{*}& E  & 32.77&$\#$  (v3K) & $-19.73\pm0.40$& $15.0\pm0.7$ & E\nl
%\hline
\enddata
\tablerefs{E = ESO+Asiago SN archive;
 a = Hamuy et al. 1996a;
 b = Phillips 1993;
 c = Vaughan et al. 1995;
 d = Lira et al. 1998;
 e = Riess et al. 1996;
 f = Turatto et al. 1996;
 g = Patat et al. 1996;
 h = Ho \& Filippenko 1995;
 i = Della Valle et al. 1998;
 l = Saha et al. 1997;
 m = Mazzali et al. 1995;
 n = Branch 1984;
 o = Cristiani et al. 1992;
 p = Wells et al. 1994;
 q = B. Leibundgut 1996, priv. comm.;
$\#$ calculated from $v3k$ from RC3 adopting H=65 km s$^{-1}$ Mpc$^{-1}$}
\tablenotetext{*}{only galactic absorption from RC3}
\end{deluxetable}
%\end{table}

\newpage

  \figcaption[snialateapjlfig1.ps]
{Optical spectra in the region 4400-5700\AA\ for four 
representative SNe~Ia. The spectrum of SN~1986G has been corrected for 
extinction assuming E(B-V)=0.6. The dotted lines show the 
multiple gaussian fit performed to measure the FWHM of the emission. 
For SN~1991bg, the dashed line is the gaussian fit of the 4660\AA\ line.}
\label{nebspectra} 

 \figcaption[snialatefwdm15.ps] 
{The value of $\Delta m_{15}(B)$ is plotted against
the FWHM of the $\sim4700$\AA\ emission feature. 
The best fit line is a weighted, least squares fit of
the data points, and has equation 
$\Delta m_{15}(B) = (1.24\pm0.03) - (0.10\pm0.01) \times (FWHM-15000)$.}
\label{FWvsDm15}

\begin{figure}
\plotfiddle{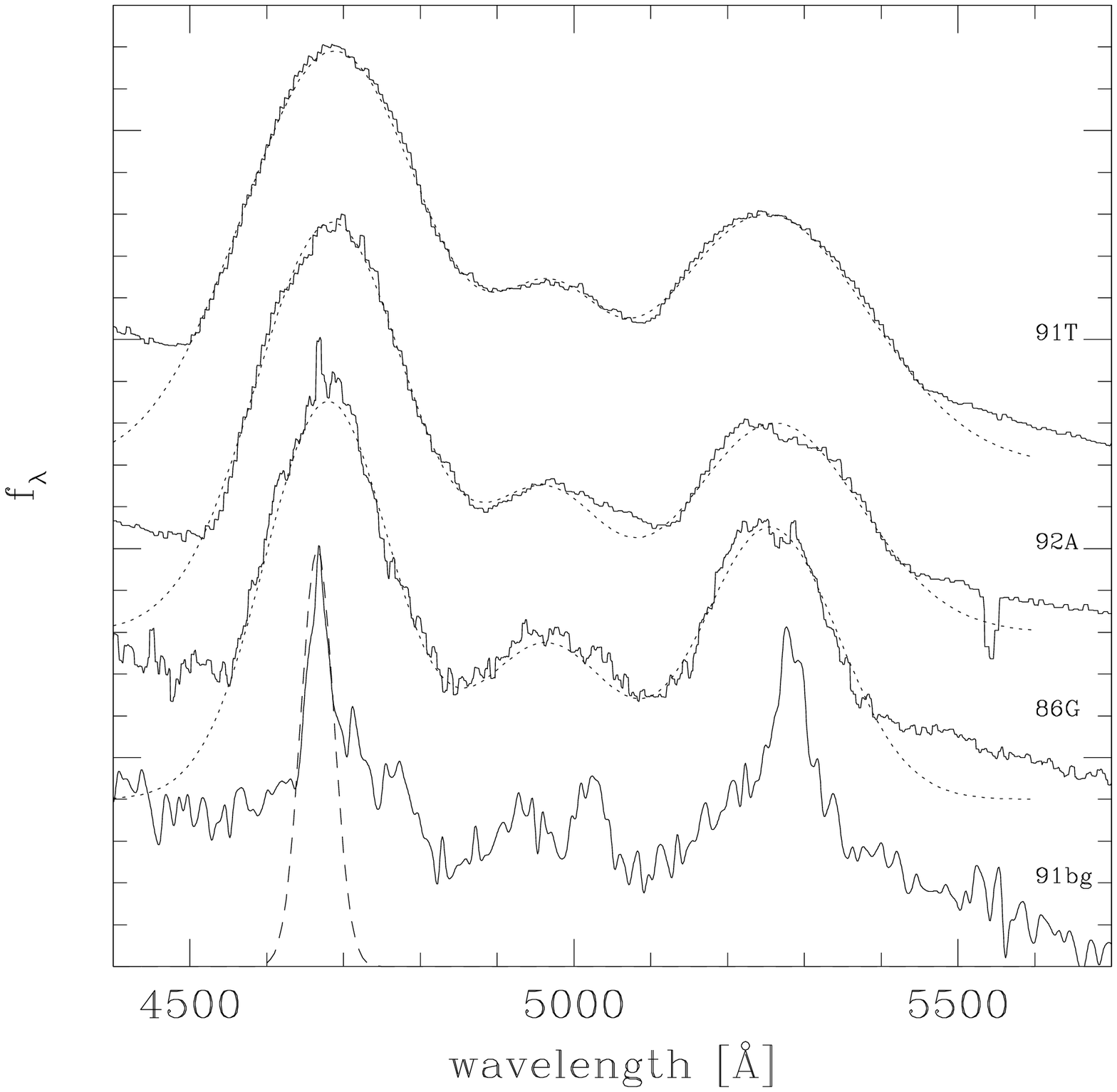}{12cm}{0}{50}{50}{-150}{50}
\end{figure} 

\begin{figure}
\plotfiddle{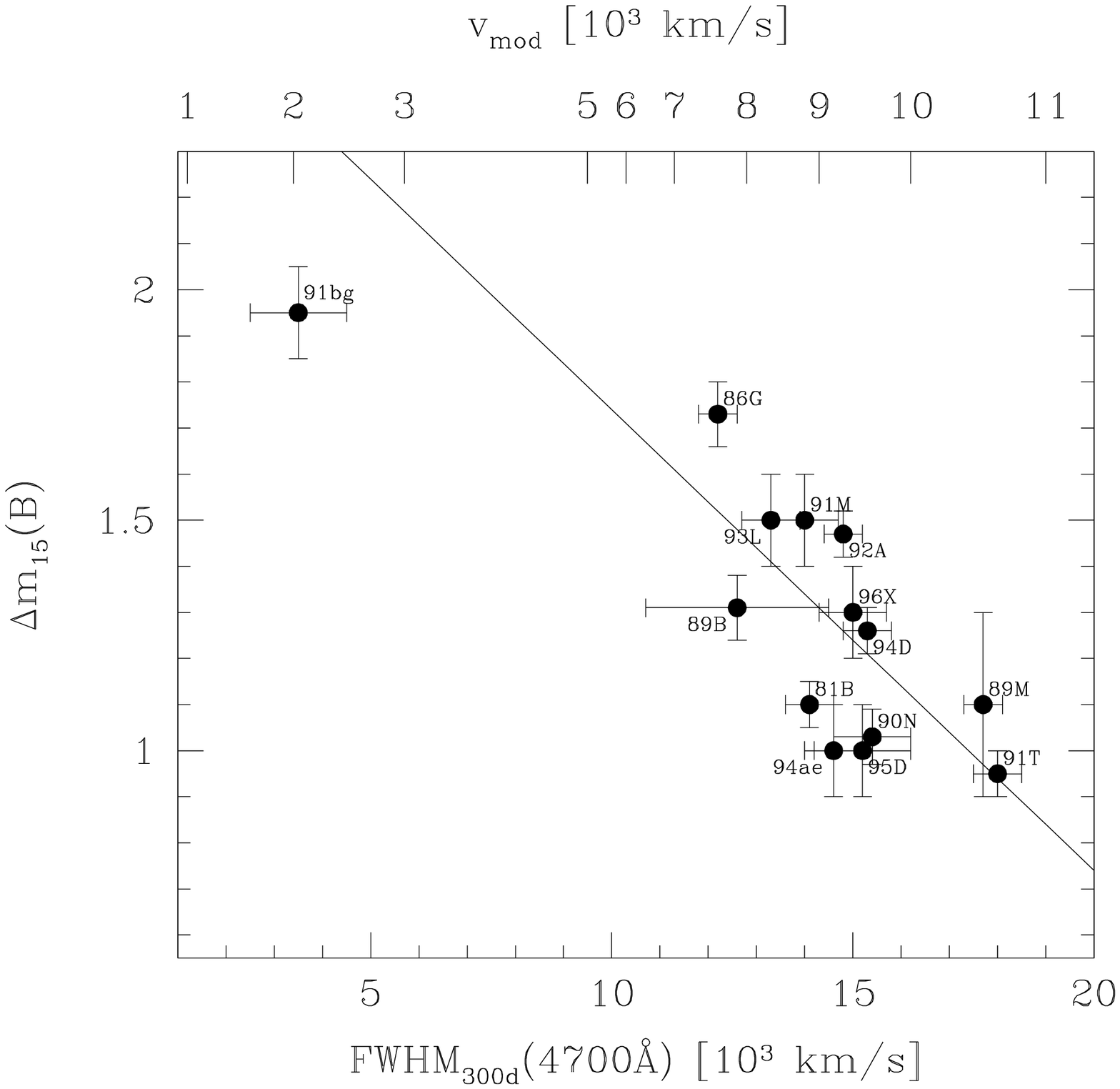}{12cm}{0}{50}{50}{-150}{50}
\end{figure} 

\end{document}